\UseRawInputEncoding
\documentclass[pra,twocolumn,10pt,superscriptaddress]{revtex4-1}  %
\usepackage{graphicx,epsfig,epstopdf}
\usepackage[dvipsnames]{xcolor}
\usepackage{amsmath}
\usepackage{subfigure}
\usepackage{mathrsfs}
\usepackage{bm}
\usepackage{graphicx}
\usepackage{times}
\usepackage{upgreek}

\begin{document}

\title{Fast quantum state transfer and entanglement for cavity-coupled many qubits via dark pathways}

\author{Yi-Xuan Wu}

\author{Zi-Yan Guan}

\author{Sai Li}

\affiliation{Guangdong Provincial Key Laboratory of Quantum Engineering and Quantum Materials,
and School of Physics\\ and Telecommunication Engineering, South China Normal University, Guangzhou 510006, China}

\author{Zheng-Yuan Xue}\email{zyxue83@163.com}
\affiliation{Guangdong Provincial Key Laboratory of Quantum Engineering and Quantum Materials,
and School of Physics\\ and Telecommunication Engineering, South China Normal University, Guangzhou 510006, China}
\affiliation{Guangdong-Hong Kong Joint Laboratory of Quantum Matter, and Frontier Research Institute for Physics,\\ South China Normal University, Guangzhou 510006, China}

\date{\today}

\begin{abstract}
Quantum state transfer (QST) and entangled state generation (ESG) are important building blocks for modern quantum information processing. To achieve these tasks, convention wisdom is to consult the quantum adiabatic evolution, which is time-consuming, and thus is of low fidelity. 
Here, using the shortcut to adiabaticity technique, we propose a general method to realize high-fidelity fast QST and ESG in a cavity-coupled many qubits system via its dark pathways, which can be further designed for high-fidelity quantum tasks with different optimization purpose. Specifically, with a proper dark pathway, QST and ESG between any two qubits can be achieved without decoupling the others, which simplifies experimental demonstrations. Meanwhile, ESG among all qubits can also be realized in a single step. In addition, our scheme can be implemented in many quantum systems, and we illustrate its implementation on superconducting quantum circuits. Therefore, we propose a powerful strategy for  selective quantum manipulation, which is promising in cavity coupled quantum systems and could find many convenient applications in  quantum information processing.
\end{abstract}

\maketitle

\section{Introduction}
Quantum computers are believed to be capable of processing some problems which are hard  for classical computer, such as factoring integers \cite{KK1} and exhaustive search \cite{KP2}, and their realization relies heavily on precise quantum control. Generally, quantum control can be regarded as finding ways of inducing quantum state transfer (QST) from an arbitrary initial quantum state to a desired target quantum state \cite{KE2}. QST is an essential element for quantum network \cite{QE1} and on-chip quantum information processing. Meanwhile, entangled state generation (ESG) are important and necessary resource in many quantum tasks \cite{book}, such as quantum teleportation \cite{KK2}, quantum dense coding \cite{KK3}, quantum cryptography \cite{KK4} and so on. Therefore, QST and ESG of arbitrary qubit with high fidelity play a very important role in scalable quantum information processing \cite{tun1, pm1, pm}.

In recent years, elementary quantum control for many qubits has been realized by using different strategies, two of which are resonant techniques \cite{QE2,QE22} and adiabatic pathway protocol \cite{ad1,ad2,ad3}. The adiabatic way is very robust against certain  errors, but it requires relatively long operational time, which will result in inevitable and unwanted information loss for quantum system without long  coherent times. Therefore, the  ¡°shortcut to adiabaticity¡± (STA)  technique \cite{STA2,STA3} is proposed to speed up the adiabatic process, which includes inverse engineering based on Lewis Riesenfeld (LR) invariant \cite{Yan}, fast-forward technique \cite{FF}, Lie algebraic methods \cite{kang} and transitionless driving \cite{STA4,Baksic,Zhou,YC,Claeys}, ect. However, they need complex  procedure, even impossible, to find the target pathways, especially for many-qubit or high-dimensional quantum systems. Besides, it usually applicable only to the global manipulation for the target quantum system. Therefore, it is highly desired to find a way of  realizing arbitrary high-fidelity quantum control in a many-qubit system.

Meanwhile, in trapped ions \cite{ion}, cavity \cite{qed1} or circuit \cite{qed2} QED systems, the bosonic quantum mode  can be used as a quantum bus to couple different qubits, known as Tavis-Cummings model \cite{MH1}, for quantum information processing, where selective QST and ESG between qubits are highly preferred. However, it is difficult to suppress the unwanted quantum state transitions for a multi-qubit system in the process of manipulating two arbitrary qubits. Conventionally, all the idle qubits need to be decoupled from the quantum bus to remove its influence, which needs additional control elements for each qubit \cite{QE55,QE5}, which will inevitably increase the complexity of the quantum circuit and thus will introduce many additional control error sources. Therefore, it is highly desired to realize QST and ESG in the multi-qubit system in a simple setup without additional control elements.

Here, we find a series of desired pathways to realize flexible QST and ESG among qubits, which possess the following distinct merits. Firstly, QST and ESG can be achieved not only with any two qubits, but also with any number of the involved qubits, without decoupling the unwanted qubits. Secondly, our proposal can be directly implemented in various cavity-coupled quantum systems \cite{ion, qed1, qed2}. Thirdly, our approach has enough flexibility for quantum system with different limitations, as it can be combined with different optimization purposes.
Besides, we illustrate our scheme  on superconducting quantum circuits with current achievable experimental technology. Through numerical simulations,  we find that faithful design of fast and robust pathways against control errors and information loss for high-fidelity quantum tasks can be obtained. Therefore, our proposal is promising in cavity-coupled quantum systems and could find many convenient applications in large-scale quantum information processing.

\section{The scheme with dark pathways}
\begin{figure}[tbp]
  \centering
  \includegraphics[width=0.9\linewidth]{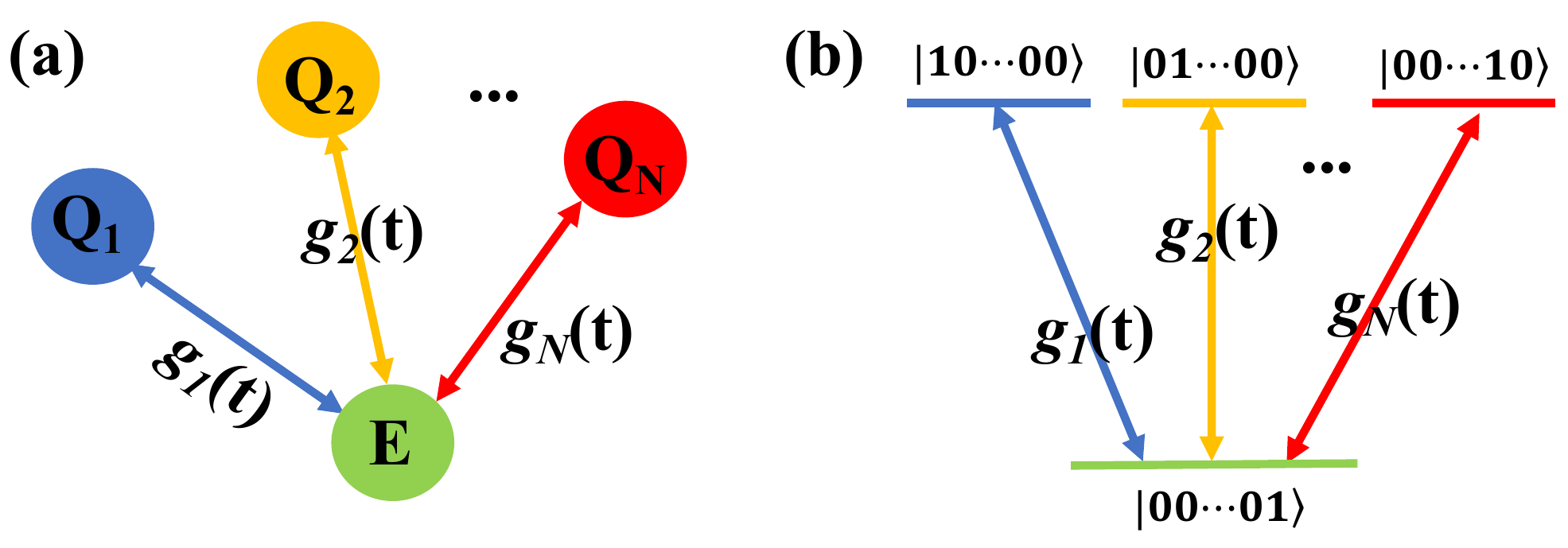}
  \caption{Illustration of the considered  quantum system. (a) $N$ qubits  resonantly coupled to an auxiliary bosonic quantum bus, with the effective coupling strength being $g_j(t)$. (b) The effective coupling structure of the  considered scenario in the single-excitation subspace.}
  \label{Figure1}
\end{figure}

Here, we consider a quantum system of many qubits that are resonantly coupled to a common  cavity, i.e., a bosonic quantum bus interacts with \emph{N} qubits \cite{CL1,CO1}. Our goal is to implement selectively quantum manipulation between any two or among more qubits, without quantum switches that decoupled the unselected qubits.
The single-excitation subspace of this coupled system is $S_1=\{ |1\rangle, |2\rangle ...|N\rangle, |a\rangle \}$,  {where $|1\rangle=|10...0\rangle$, $|N\rangle=|0...010\rangle$ labeling the product states of \emph{N} qubits and $|a\rangle=|0...001\rangle$ labeling the quantum bus with $| Q_1 Q_2...Q_N Q_a\rangle\equiv|Q_1\rangle\otimes|Q_2\rangle\otimes...|Q_N\rangle\otimes|Q_a\rangle$}, as shown in Fig. \ref{Figure1}. Assuming $\hbar=1$, hereafter, in the interaction picture, the Hamiltonian of the coupled system is \cite{MH1}
\begin{equation}
\label{E1}
H_N(t)=\sum_{j=1}^N g_j(t)|j \rangle \langle a|+ \text{H.c.},
\end{equation}
where $ g_j(t)$ is the time-dependent tunable effective coupling strength between the $j$th qubit and the   quantum bus.

Then, we illustrate how to find a target dark pathway, denoted by $|\psi(t)\rangle$, aiming at multi-qubit QST and ESG. This pathway needs to satisfy two conditions: the first one is the natural normalization principle, and the second is that its expectation value is zero, i.e.,
\begin{equation}
\label{E2}
 \langle\psi(t)|H_N(t)|\psi(t)\rangle=0.
\end{equation}
That is to say, during the whole evolution process, there is no dynamical phase accumulated. Thus, we define the state $|\psi(t)\rangle$ with zero expectation value as dark pathway, similar to dark state with zero eigenenergy.


We now focus on the QST and ESG between two arbitrary qubits from the \emph{N} qubits system in the given time $T$. For example, the state of the quantum system transfer from $|m\rangle$ to $|n\rangle$ or $(|m\rangle+|n\rangle)/\sqrt{2}$ through the dark pathways. In order to satisfy two conditions of the dark pathways, we can construct one of the dark pathways as
\begin{eqnarray} \label{E11}
|\psi(t)\rangle_2&= &\cos\gamma_1\cos\gamma_2\cos\gamma_3|m\rangle-\cos\gamma_1\cos\gamma_2\sin\gamma_3|n\rangle\notag\\
&-&\frac{1}{\sqrt{N-2}}\sum_{j\neq{m,n}}\cos\gamma_1\sin\gamma_2|j\rangle-i\sin\gamma_1|a\rangle,
\end{eqnarray}
where $\gamma_{1,2,3}$ are the auxiliary parameters. It is worth noting that this is only a general pathway rather than a special one. Of course, we can also choose other pathways just they can meet the conditions. Substituting it into Schr\"{o}dinger equation, we can obtain the tunable coupling strength as
\begin{equation}
\label{E12} \small{
g_j^{(2)}(t)=\left\{
\begin{aligned}
   \dot\gamma_1\cos\gamma_2\cos\gamma_3&+\dot\gamma_2\cot\gamma_1\sin\gamma_2\cos\gamma_3\\
           &+\dot\gamma_3\cot\gamma_1\cos\gamma_2\sin\gamma_3, \quad j=m\\
  \dot\gamma_3\cot\gamma_1\cos\gamma_2&\cos\gamma_3-\dot\gamma_1\cos\gamma_2\sin\gamma_3\\
           &-\dot\gamma_2\cot\gamma_1\sin\gamma_2\sin\gamma_3,  \quad j=n\\
   (\dot\gamma_2\cot\gamma_1\cos\gamma_2&-\dot\gamma_1\sin\gamma_2)/\sqrt{N-2}, \quad j\neq m,n \\
\end{aligned}
\right.}
\end{equation}
Therefore, the implementation of QST can be achieved via this dark pathway. According to the initial state, target state and Eq. (\ref{E11}), we can get the boundary conditions as
\begin{equation}
\label{E13}
    \gamma_{1, 2}(0)=0,  \gamma_{1, 2}(T)=0,   \gamma_3(0)=0,  \gamma_3(T)=\theta,
\end{equation}
where $\theta_S=\pi/2$ and $\theta_E=\pi/4$ for the fast QST and ESG, respectively. For simplification, we can set $H(0)=0$, $H(T )=0$ to meet the experimental restriction. Then, $\gamma_{1, 2, 3}$ can be deduced as
\begin{equation}
\label{E14}
   \dot\gamma_{1,2,3}(0)= \dot\gamma_{1,2,3}(T)=0,
\end{equation}
According to Eq. (\ref{E13}) and Eq. (\ref{E14}), we can get different forms of auxiliary parameters in different ways, such as Taylor series expansion \cite{IM4}, polynomial expansion, GRAPE algorithm \cite{IM2} that just adds some constraints on the basis of Eq. (\ref{E13}) and Eq. (\ref{E14}), and so on. What we use here is a relatively simple binomial expansion method,
thus the simple form of the auxiliary parameters can be chosen as \cite{IM3}
\begin{equation}
\label{E15}
\begin{aligned}
   \gamma_1(t)&=\frac{A}{(T/2)^4}t^2(t-T)^2, \\
   \gamma_2(t)&=1-\cos\gamma_1, \\
   \gamma_3(t)&=\frac{-20\theta t^7}{T^7}+\frac{70\theta t^6}{T^6}-\frac{84\theta t^5}{T^5}+\frac{35\theta t^4}{T^4},
\end{aligned}
\end{equation}
where $A$ is a tunable constant parameter. Note that high-fidelity QST and ESG can be optimized by different options of the parameter $A$  for different purposes, for various quantum systems with different limitations  \cite{MM1}, including reducing population of the auxiliary states \cite{IM3}, improving the robustness against system errors \cite{IM4}, shortening evolution time against decoherence and so on. Therefore, our scheme is applicable for quantum systems with different constraints.


We next move to ESG for all qubits in this system, in the case of controlling an initial state $|m\rangle$ evolves to $\sum^N_{j=1}{|j\rangle}/\sqrt{N}$. In order to realize ESG for all qubits, we construct the dark pathway as
\begin{eqnarray} \label{E16}
|\psi(t)\rangle_N&=&\cos\gamma_1'\cos\gamma_2'|m\rangle
-\frac{1}{\sqrt{N-1}} \sum_{j \neq m} \cos\gamma_1'\sin\gamma_2'|j\rangle \notag \\
&&-i\sin\gamma_1'|a\rangle,
\end{eqnarray}
and the tunable coupling strengths as
\begin{equation}
\label{E17}
g_j^{(N)}(t)=\left\{
\begin{aligned}
    \dot\gamma_2'\cot\gamma_1'\sin\gamma_2'+\dot\gamma_1'\cos\gamma_2', \quad& j=m;\\
    \frac{1}{\sqrt{N-1}}(\dot\gamma_2' \cot\gamma_1' \cos\gamma_2' - \dot\gamma_1' \sin\gamma_2') , \quad& j\neq m.\\
\end{aligned}
\right.
\end{equation}
According to the initial state, target state and Eq. (\ref{E16}), we can get the same boundary conditions as
\begin{equation}
\label{E18}
    \gamma_1'(0)=0,\quad \gamma_1'(T')=0,\quad  \gamma_2'(0)=0,\quad   \gamma_2'(T')=\theta'
\end{equation}
with $\theta'=\arccos\sqrt{1/N}$, and
\begin{equation}
\label{E19}
\begin{aligned}
   \gamma_1'(t)&=\frac{A'}{(T'/2)^4}t^2(t-T')^2, \\
   \gamma_2'(t)&=\frac{-20\theta' t^7}{T'^7}+\frac{70\theta' t^6}{T'^6}-\frac{84\theta' t^5}{T'^5}+\frac{35\theta' t^4}{T'^4},
\end{aligned}
\end{equation}
where $A'$ is a tunable constant parameter, which can be similarly used to design for high-fidelity purpose in quantum systems with different optimization purpose.

\begin{figure}[tb]
	\centering
	\includegraphics[width=0.85\linewidth]{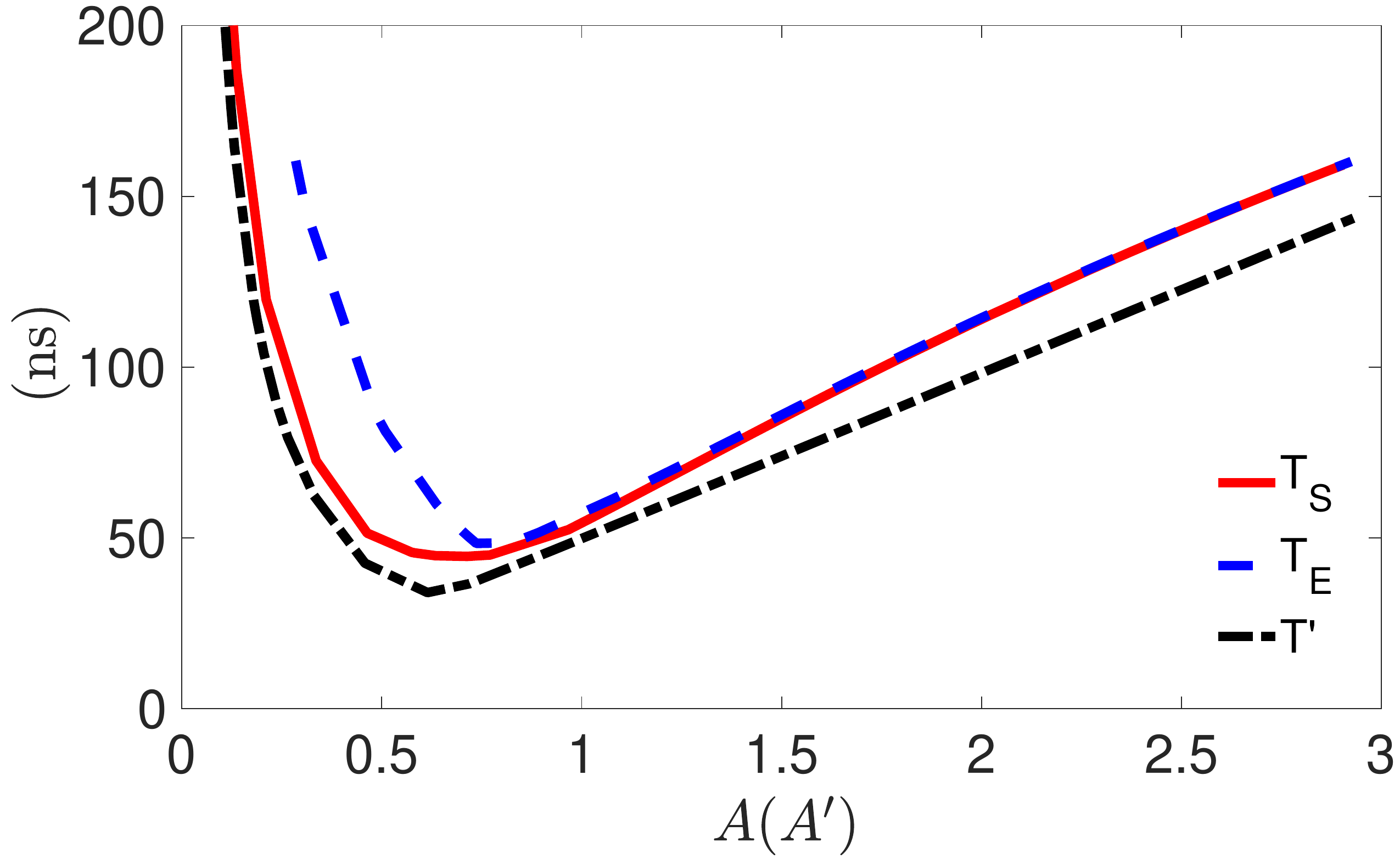}
\caption{ The evolution time as a function of the parameter $A(A')$  for the QST ($T_S$), ESG of two qubits ($T_E$) and ESG for all qubits ($T'$), respectively. The optimal $A(A')$ is chosen when the evolution time is the minimum value. }\label{Figure2}
\end{figure}

\begin{figure}[tb]
	\centering
	\includegraphics[width=0.9\linewidth]{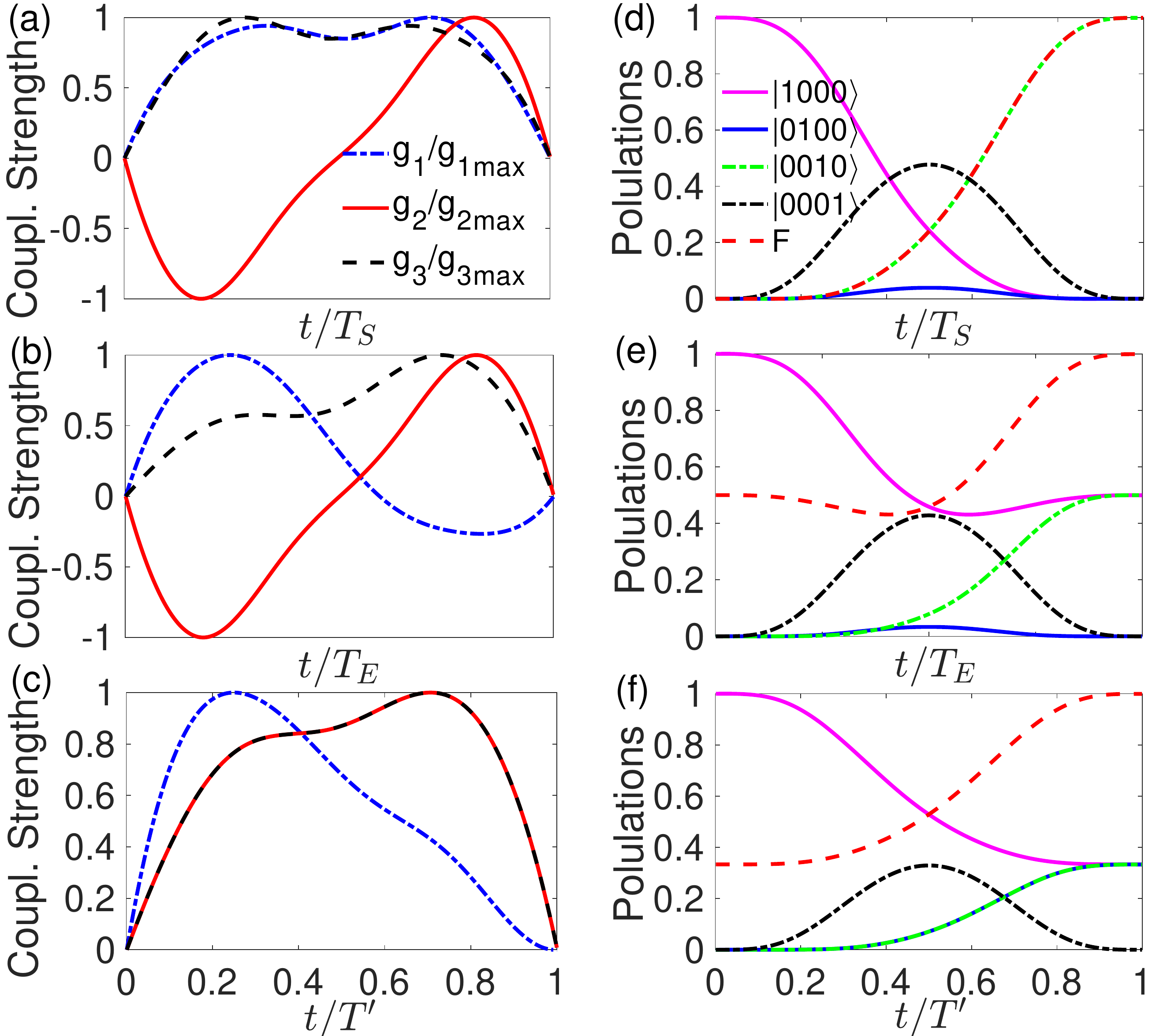}
	\caption{Numerical simulation of the \emph{N}=3 case. Coupling strength of (a) QST, (b) ESG of two qubits and (c) ESG for all qubits determined by $\gamma_{1,2,3}$ ($\gamma'_{1,2}$).   Populations and state fidelity of (d) QST and (e) ESG from the initial state $|1000\rangle$ to the final states $|0010\rangle$ and $(|1000\rangle+|0010\rangle)/\sqrt2$, respectively. (f) ESG for all the qubits, i.e., in the case of quantum state evolution from $|1000\rangle$ to $(|1000\rangle+|0100\rangle+|0010\rangle)/\sqrt{3}$. }
\label{Figure3}
\end{figure}

\section{Examples with $N=3$}

In this section, we  detail our statement with a specific example, the $N=3$ case in the last Section.  According to Eq. (\ref{E1}), the Hamiltonian of the coupled system in $N=3$ case is
\begin{eqnarray}
\label{E111}
H_3(t)&=&g_1(t)|1000 \rangle \langle 0001|+g_2(t)|0100 \rangle \langle 0001|   \notag\\
&&+g_3(t)|0010 \rangle \langle 0001|+ \text{H.c.}.
\end{eqnarray}
First, we illustrate the QST and ESG of any two qubits for $N=3$ case, i.e., from $|1000\rangle$ to $|0010\rangle$ or $(|1000\rangle+|0010\rangle)/\sqrt{2}$. In this case, the dark pathway in Eq. (\ref{E11}) is chosen as
\begin{equation}
\begin{aligned}
\label{E26}
|\psi(t)\rangle_2= &\cos\gamma_1\cos\gamma_2\cos\gamma_3|1000\rangle -\cos\gamma_1\sin\gamma_2|0100\rangle \\
& -\cos\gamma_1\cos\gamma_2\sin\gamma_3|0010\rangle -i\sin\gamma_1|0001\rangle,\\
\end{aligned}
\end{equation}
and the effective coupling strengths in Eq. (\ref{E12}) is chosen as
\begin{equation}
\label{E27}
\begin{aligned}
   g_1^{(2)}(t)=&\dot\gamma_1\cos\gamma_2\cos\gamma_3+\dot\gamma_2\cot\gamma_1\sin\gamma_2\cos\gamma_3\\
           &+\dot\gamma_3\cot\gamma_1\cos\gamma_2\sin\gamma_3, \\
  g_2^{(2)}(t)=&\dot\gamma_3\cot\gamma_1\cos\gamma_2\cos\gamma_3-\dot\gamma_1\cos\gamma_2\sin\gamma_3\\
           &-\dot\gamma_2\cot\gamma_1\sin\gamma_2\sin\gamma_3,  \\
   g_3^{(2)}(t)=&\dot\gamma_2\cot\gamma_1\cos\gamma_2-\dot\gamma_1\sin\gamma_2, \\
\end{aligned}
\end{equation}
where $\gamma_{1,2,3}$ are in the same form as Eq. (\ref{E15}).
To achieve the fast QST with high fidelity, it is better to set the time as short as possible. After setting the condition of the maximum effective coupling strength  max$\{g_j(t)\}=g_{j\text{max}}$, as shown in Fig. \ref{Figure2}, we plot the evolution time as a function of the parameter $A$ and find the minimum time point. Here, we find the optimal $A_S=0.7365$ corresponds to the shortest time  $T_S\simeq3/g$ for QST, with max$\{g_{j\text{max}}\}=g$, then the dark pathway $|\psi(t)\rangle_2 $ and the slope of the coupling strength $ g_j^{(2)}(t)$, as shown in Fig. \ref{Figure3} (a), can be determined. Similarly, the optimal value $A_E=0.7138$ and $T_E\simeq2.8/g$ for ESG of two qubits can be found and the coupling strength $ g_j^{(2)}(t)$ is shown in Fig. \ref{Figure3} (b).

To evaluate this process, we numerically simulate the quantum evolution using the Lindblad master equation of
\begin{equation}
\begin{split}
\label{E28}
\dot{\rho} =&i[\rho, H_3(t)]+\sum_{k=1,2,3}(\Gamma_{1k} \mathcal{L}( \sigma^{-}_{1k})+\Gamma_{2k} \mathcal{L}( \sigma^{z}_{2k}))\\
&+\Gamma_{1a} \mathcal{L}( \sigma^{-}_{1a})+\Gamma_{2a} \mathcal{L}( \sigma^{z}_{2a}),
\end{split}
\end{equation}
where $\rho$ is the density matrix of the considered system, $\mathcal{L}(\mathcal{D})= \mathcal{D} \rho \mathcal{D}^{\dagger} -\mathcal{D}^{\dagger} \mathcal{D} \rho/2-\rho\mathcal{D}^{\dagger} \mathcal{D}/2$ is the Lindblad operator of $\mathcal{D}$ with $\sigma^{-}_{1k}=|0\rangle_k \langle 1|$, $\sigma^{-}_{1a}=|0\rangle_a \langle 1|$ and $\sigma^{z}_{2k}=|0\rangle_k \langle 0|-|1\rangle_k \langle 1|$, $\sigma^{z}_{2a}=|0\rangle_a \langle 0|-|1\rangle_a \langle 1|$, and $\Gamma_{1k}$, $\Gamma_{2k}$, $\Gamma_{1a}$ and $\Gamma_{2a}$ are the decay and dephasing rates of qubits. We consider the simple case of $\Gamma_{1k}=\Gamma_{2k}=\Gamma_{1a}=\Gamma_{2a}=\Gamma = g/2000$, which is easily accessible with experimental technologies, even for superconducting qubits \cite{GE11}. In order to have an insight into the detailed population changes, we numerically simulate the preparation of state transfer from $|1000\rangle$ to $|0010\rangle$, as shown in Fig. \ref{Figure3} (d), where we obtain a fidelity of $99.89\%$. The ESG process is evaluated  in Fig. \ref{Figure3} (e), and a fidelity of $99.91\%$ is obtained. Here, the infidelity is mainly due to the  decoherence effect under the constraint of the maximum coupling strength.


We now move to the case that ESG of all the 3 qubits, i.e., the case of state from $|1000\rangle$ to $(|1000\rangle+|0100\rangle+|0010\rangle)/\sqrt{3}$. In this case, we can construct the dark pathway in Eq. (\ref{E16}) with $N=3$ as
\begin{eqnarray}
\label{E29}
|\psi(t)\rangle_N&=&\cos\gamma_1'\cos\gamma_2'|1000\rangle-\frac{1}{\sqrt{2}}\cos\gamma_1'\sin\gamma_2'|0100\rangle  \notag\\
&&-\frac{1}{\sqrt{2}}\cos\gamma_1'\sin\gamma_2'|0010\rangle-i\sin\gamma_1'|0001\rangle,
\end{eqnarray}
and the effective coupling strengths in  Eq. (\ref{E17}) as
\begin{equation}
\label{E30}
\begin{aligned}
    g_1^{(N)}(t)=&\dot\gamma_2'\cot\gamma_1'\sin\gamma_2'+\dot\gamma_1'\cos\gamma_2', \\
    g_2^{(N)}(t)=&\frac{1}{\sqrt{2}}(\dot\gamma_2' \cot\gamma_1' \cos\gamma_2' - \dot\gamma_1' \sin\gamma_2') ,\\
    g_3^{(N)}(t)=&\frac{1}{\sqrt{2}}(\dot\gamma_2' \cot\gamma_1' \cos\gamma_2' - \dot\gamma_1' \sin\gamma_2') ,
\end{aligned}
\end{equation}
where $\gamma_1'$ and $\gamma_2'$ are as in the same form as Eq. (\ref{E19}) with $\theta'=\arccos \sqrt{1/3}$. Similarly,  to achieve ESG for all qubits with high fidelity, we can find the optimal value $A'=0.6143$ and  $T'\simeq2\pi/(3g)$. The coupling strength of ESG for all qubits is shown in Fig. \ref{Figure3}(c) and the generation process is simulated as shown in Fig. \ref{Figure3}(f), with a fidelity of $99.94\%$.

\begin{figure}[tb]
	\centering
	\includegraphics[width=1\linewidth]{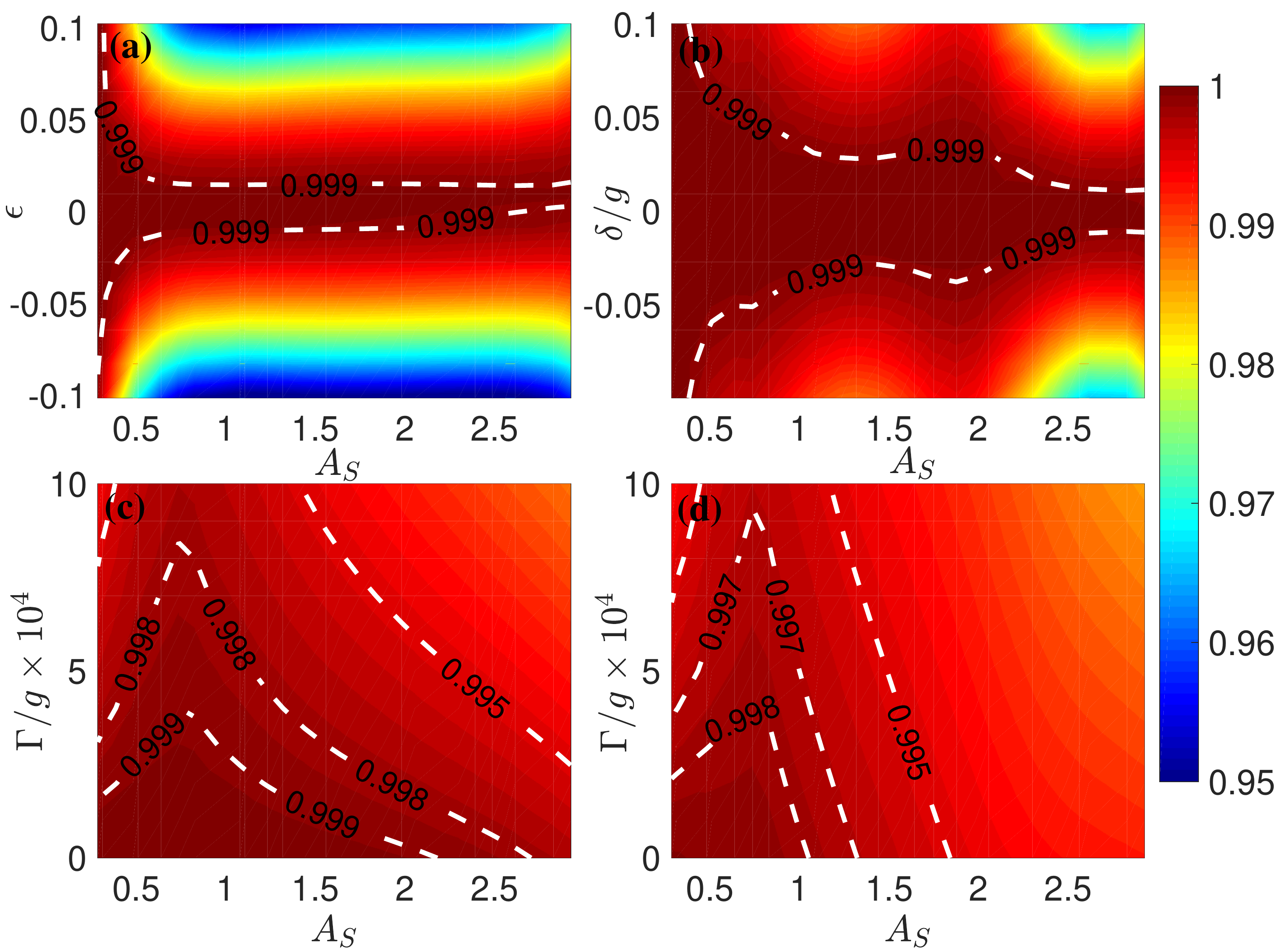}
\caption{Robustness of our scheme. Without considering the decoherence effect, state fidelity (a) as a function of the X error and $A_S$, (b) as a function of the $\sigma_z$ error and $A_S$. (c) All qubits with the uniform decoherence rate of $\Gamma_{1k}=\Gamma_{2k}=\Gamma_{1a}=\Gamma_{2a}\in  [0, g/1000]$. (d) The decoherence rate of auxiliary qubit is set to be $g/1000$ while other qubits within the range $[0, g/1000]$.} \label{Figure4}
\end{figure}

Furthermore, we take the QST of two qubits as an example to show the robustness of our scheme against different errors, with parameters $A_S$ and $T_S$ are determined shown in Fig. \ref{Figure2}. Taking the operational control error caused by the deviation $\epsilon$ of driving amplitude, i.e., X error, the Hamiltonian in Eq. (\ref{E1}) turns to $H'_N(t)=(1+\epsilon)H_N(t)$. Considering the error range $-0.1\leq\epsilon\leq0.1$, we plot the state fidelities as a function of the deviation $\epsilon$ of driving amplitude and parameter $A_S$ without decoherence as shown in Fig. \ref{Figure4} (a). Considering another operational control error caused by the $\sigma_z$ error, i.e., randomized qubit-frequency-drift-induced error, which is in the form of $\delta\sum_{j=1}^N |j \rangle \langle j|$ with $\delta$ being the drift quantity. In the presence of  frequency drifts  for all the involved qubits, the Hamiltonian in Eq. (\ref{E1}) will change to
$H'_{0N}(t)=H_N(t)+\delta\sum_{j=1}^N |j \rangle \langle j|$.
Setting $\delta\in [-g/10, g/10]$, as shown in Fig. \ref{Figure4} (b), we simulate the fidelity as a function of $A_S$ and $\delta$ without decoherence. We found that, with the decrease of $A_S$, the needed time $T_S$ for the process will become  longer and the robustness against noise is better.  From Fig. \ref{Figure4} (a) and Fig. \ref{Figure4} (b), when $A_S$ is smaller, the evolution process is more robust against both X and $\sigma_z$ errors. But from Fig. \ref{Figure2}, when $A_S$ is smaller, the evolution time $T_S$ is longer. This is natural, in the limiting case, with much smaller $A_S$, this process will reduce to the well-known adiabatic process, which possesses very strong robustness. Therefore, there is a balance point between the robustness and evolution time in realistic physical quantum systems. Also, the decoherence caused by environment is the main factor for state infidelity. We plot the state fidelity as a function of the parameter $A_S$ and the rate of decoherence $\Gamma$. First, setting $\Gamma_{1k}=\Gamma_{2k}=\Gamma_{1a}=\Gamma_{2a}\in  [0, g/1000]$, as shown in Fig. \ref{Figure4}(c). Then, the decoherence rate of the auxiliary qubit is set $\Gamma_{1a}=\Gamma_{2a}= g/1000$  while $\Gamma_{1k}=\Gamma_{2k}\in  [0, g/1000]$, as shown in Fig. \ref{Figure4}(d). Obviously, in the presence of the decoherence, the minimum time point possesses the strongest robustness. Also, comparing Fig. \ref{Figure4}(c) and Fig. \ref{Figure4}(d), for the same qubits decoherence, it is clear that the fidelity in Fig. \ref{Figure4}(d) decreases faster than Fig. \ref{Figure4}(c) with the increase of $A_S$. This  is due to the fact that larger $A_S$ leads to more  auxiliary state population, and thus large decoherence of the auxiliary devices will lead to lower  fidelity of the process.

\section{Physical implementation with transmons}

In this section, we propose a scheme on superconducting quantum circuits to demonstrate our protocol in detail. What we consider is three superconducting transmon qubits are coupled to a common transmission-line resonator \cite{GE1,GE2,GE3}, and the transmons with the lowest two levels $|0\rangle$ and $|1\rangle$ serving as qubit states, where higher energy levels of the transmons are not considered due to the fact that only quantum dynamics within the single-excitation subspace is involved here. We label three transmons as $A$, $B$ and $C$ with frequencies $\omega_{A,B,C}$, and the superconducting transmission line resonator $E$ with frequency $\omega_E$. Assuming $\hbar=1$, hereafter, the Hamiltonian of the coupled system can be described as
\begin{eqnarray}
\label{E20}
H_S(t)&=&\sum_{j\in \{A,B,C\}} \{[\omega_j+f(\varepsilon_j(t))]|1\rangle_j\langle1|\}+\omega_E|1\rangle_E\langle1|\notag\\
&&+\sum_{j=A,B,C}\Omega_j\sigma^x_j (|0\rangle_E\langle1|+\text{H.c.}),
\end{eqnarray}
where $\Omega_j$ is the coupling strength for transmons $A,B,C$ to $E$, and $f(\varepsilon_j(t))$ is the nonlinear frequency response to the modulation pulse on the transmon, which can be determined experimentally by the longitudinal field $\varepsilon_j(t)=f^{-1}(\dot{F}_j(t))$ with $F_j(t)=\eta_j(t)\sin(\nu_j t)$  \cite{WL2,pm2}. Moving to the rotating frame defined by $U=U_1\times U_2$, where \cite{IM1}
\begin{equation}
\label{E21}
\begin{split}
&U_1=\exp\left[-i\sum_{j\in \{A,B,C\}}(\omega_{j}|1\rangle_j\langle1|+\omega_E|1\rangle_E\langle1|)\right],\\
&U_2=\exp\left[i\sum_{j\in \{A,B,C\}}F_j(t)|1\rangle_j\langle1|\right],
\end{split}
\end{equation}
the transformed Hamiltonian is given by
\begin{equation}
\label{E22}
\begin{split}
H_{ST}(t)=&U^\dagger H_S(t) U+i \frac{dU^\dagger}{dt}U\\
=&\sum_{j=A,B,C}\Omega_j[|0\rangle_j\langle1| e^{-i\omega_j t}e^{iF_j(t)}+\text{H.c.}]\\
&\otimes[|0\rangle_E\langle1| e^{-i\omega_Et}+\text{H.c.}],
\end{split}
\end{equation}
in the single-excitation subspace spanned by $\{|1000\rangle,|0100\rangle,|0010\rangle,|0001\rangle\}$, where $|ABCE\rangle \equiv|A\rangle\otimes|B\rangle\otimes|C\rangle\otimes|E\rangle$ labels the product states of three transmons and the resonator.

\begin{figure}[tb]
	\centering
	\includegraphics[width=0.9\linewidth]{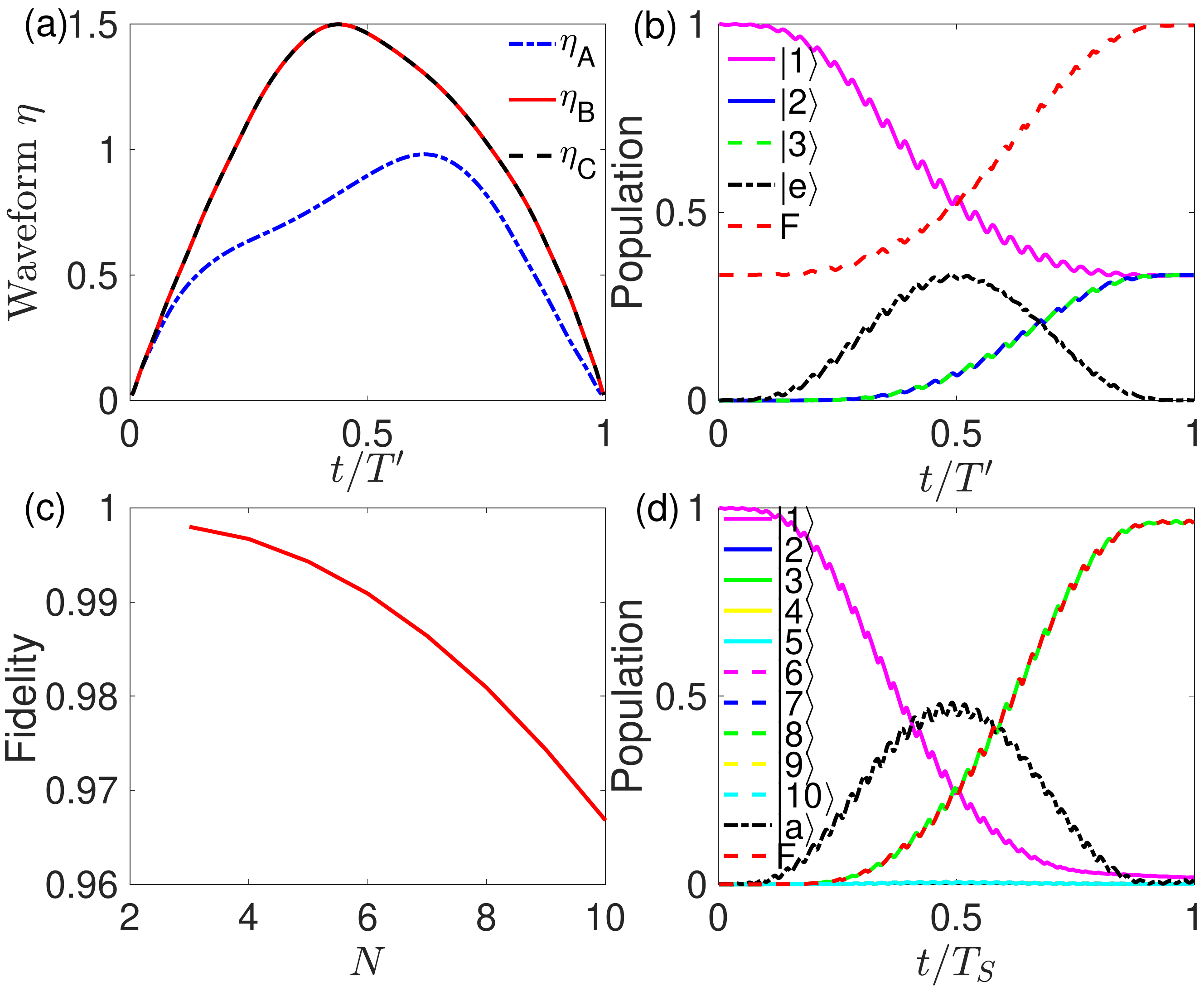}
	\caption{ Numerical simulation of (a) the $\eta_j$ waveform of ESG for all qubits in the $N=3$ case, populations and state fidelity of (b) ESG for all qubits in $N=3$ case and (d) the QST in $N=10$ case. (c) State fidelities of the QST as a function of the number $N$ of qubits. }\label{Figure5}
\end{figure}

After neglecting the high order oscillating terms, the Hamiltonian can be written as
\begin{equation}\label{E23}
\begin{split}
H_I(t)=&\Omega_A|1000\rangle\langle0001| e^{i\Delta_At-iF_A(t)}+\\
       &\Omega_B|0100\rangle\langle0001| e^{i\Delta_Bt-iF_B(t)}+\\
       &\Omega_C|0010\rangle\langle0001| e^{i\Delta_Ct-iF_C(t)}+H.c.,
\end{split}
\end{equation}
where $\Delta_j=\omega_j-\omega_E$. Using the Jacobi-Anger identity of
$e^{iF_j(t)}=\sum_{m=- \infty}^ \infty i^m J_m(\eta_j(t))\exp[im\nu_j t ]$
with $J_m(\eta_j(t))$ being the $m$th Bessel function of the first kind and considering the resonant interaction case $\Delta_j=\nu_j$, the effective Hamiltonian is simplified as
\begin{equation}
\begin{split}
\label{E25}
H_{eff}(t)=&g_A(t)|1000\rangle\langle0001|+g_B(t)|0100\rangle\langle0001|\\
        &+g_C(t)|0010\rangle\langle0001|+\text{H.c.},
\end{split}
\end{equation}
where the effective time-modulation coupling strength $g_j(t)=\Omega_jJ_1(\eta_j(t))$. We can use the effective Hamiltonian $H_{eff}(t)$  by conveniently tuning $g_j(t)$ to realize QST and ESG, which adopts a way of parametric modulation of qubit frequency to realize tunable qubit coupling strengths.  Experimentally,  each qubit-cavity coupling strength  can be tunable in a wide range via adjusting  $\eta_j(t)$ of the longitudinal driving field \cite{pm2}.
Considering the experiment realization of QST and ESG, $\eta_j (t)$ is the important manipulated parameter rather than $g_j(t)$. Therefore, we take ESG for all qubits as an example, giving the time-dependent waveforms of $\eta_j (t)$ according to the waveform of $g_j(t)$ by inversely solving the Bessel function of the first kind as shown in Fig. \ref{Figure5} (a).

We proceed to illustrate QST and ESG of any two
qubits in superconducting quantum circuits in the case of minimum evolution times as examples. We set the frequency of the longitudinal field $\nu_j$ equal to the corresponding frequency difference $\Delta_j$ as $\nu_{A,B,C}=\Delta_{A,B,C}=2\pi\times800$ MHz to induce time-modulation resonant interaction in the single-excitation subspace. And the coupling strength for transmons $A,B,C$ to $E$ can be set as $\Omega_{A,B,C}=\Omega=2\pi\times17$ MHz \cite{WL2}. In the case of QST and ESG of any two qubits for $N=3$, we can find the optimal value as $A_S=0.7365$ and $T_S=48.4$ ns, $A_E=0.7138$ and $T_E=44.6$ ns, respectively. And the dark pathway is chosen as Eq. (\ref{E26}) and the effective coupling strengths also can be chosen according to Eq. (\ref{E27}). With the decoherence rate $\Gamma=2\pi\times5$ kHz, corresponds to $\Gamma/\Omega\approx 3\times 10^{-4}$ and well within the range set in the above Section,  through numerical simulation, we can get the QST from $|1000\rangle$ to $|0010\rangle$ with a fidelity of $99.63\%$, and ESG with a fidelity of $99.12\%$, where the infidelity is caused by the longitudinal field and decoherence.

We now move to the case that ESG of all the 3 qubits, i.e., the case of state from $|1000\rangle$ to $(|1000\rangle+|0100\rangle+|0010\rangle)/\sqrt{3}$. In this case, we can construct the dark pathway as in Eq. (\ref{E29}) and the effective coupling strengths according to Eq. (\ref{E30}). In order to make the evolution time to be the shortest, we can similarly find the optimal value as $A'=0.6143$ and $T'=34$ ns, as shown in Fig. \ref{Figure5}(b), the  simulation of  the ESG process gives a fidelity of $99.68\%$.
Then, considering the influence of $N$ (the number of qubits) on fidelity. Here, we take the QST of any two qubits as an example, simulating the fidelity as a function of $N$. Clearly, we can get the decreasing curves, i.e., with the increase of $N$, the fidelity will show a downward trend, as shown on Fig. \ref{Figure5} (c). Here, we present the simulation result for the  QST between any two qubits in the $N=10$ case, as showed in Fig. \ref{Figure5} (d), and we obtain a fidelity of $96.68\%$.

\section{CONCLUSION}
In conclusion, we have proposed a scheme to achieve fast QST and ESG between any two qubits in a multi-qubit scenario with high-fidelity via the properly designed dark pathways. Our scheme can be achieved  without additional operations to turn off the unrelated qubits, which is preferred in practical quantum systems, as   selective quantum manipulation is  inevitable and difficult  in general quantum information processing tasks. Besides, single-step ESG for all qubits can also be generated in this setup. What's more, special dark pathways can be designed for high-fidelity purpose in quantum systems with different optimization purposes, making our scheme be compatible with various pulse-shaping techniques. In addition, our scheme can be directly implemented  in many multi-qubit systems which are promising candidates for the physical implementation of quantum computers. Therefore, our scheme may find many convenient applications in large-scale quantum information processing tasks.


\acknowledgements
This work was supported by the Key-Area Research and Development Program of GuangDong Province (No. 2018B030326001), the National Natural Science Foundation of China (No. 11874156),  and the Science and Technology Program of Guangzhou (No. 2019050001).

\end{document}